\documentclass[%
 reprint,
 amsmath,amssymb,
 aps,
]{revtex4-2}

\usepackage{graphicx}
\usepackage{dcolumn}
\usepackage{bm}

\usepackage{caption}
\usepackage{subcaption}
\usepackage{xcolor}
\usepackage{mathrsfs}

\newcommand{\supp}{{\rm supp}}
\newcommand{\eps}{\varepsilon}
\newcommand{\bz}{\mathbf{z}}
\newcommand{\bi}{{\mathbf i}}

\newcommand{\bj}{{\mathbf j}}
\newcommand{\bX}{{\mathbf X}}
\newcommand{\bY}{{\mathbf Y}}

\newcommand{\bx}{\mathbf{x}}
\newcommand{\by}{\mathbf{y}}
\newcommand{\bk}{\mathbf{k}}
\newcommand{\tB}{\textrm{B}}
\newcommand{\bzero}{\mathbf{0}}
\newcommand{\mO}{\mathcal{O}}
\newcommand{\blue}[1]{{\color{black} #1}}
\newtheorem{rmk}{Remark}

\newcommand\mo{
	\mathchoice
	{{\scriptstyle\mathcal{O}}}
	{{\scriptstyle\mathcal{O}}}
	{{\scriptscriptstyle\mathcal{O}}}
	{\scalebox{.7}{$\scriptscriptstyle\mathcal{O}$}}
}

\begin{document}

\preprint{APS/123-QED}

\title[A Counterexample to the L\'evy Flight Foraging Hypothesis]{A Counterexample to the L\'evy Flight Foraging Hypothesis in the Narrow Capture Framework}

\author{Justin C. Tzou}
 \email{tzou.justin@gmail.com}
 \affiliation{School of Mathematical and Physical Sciences, Macquarie University, Sydney, NSW,}
\author{Leo Tzou}%
 \email{leo.tzou@gmail.com}
\affiliation{Korteweg-de Vries Institute for Mathematics, University of Amsterdam, Amsterdam, Netherlands}%

%
%

\date{\today}

\begin{abstract}
The L\'evy flight foraging hypothesis asserts that biological organisms have evolved to employ (truncated) L\'evy flight searches due to such strategies being more efficient than those based on Brownian motion. However, we provide here a concrete two-dimensional counterexample in which Brownian search is more efficient. In fact, we show that the efficiency of L\'evy searches worsens the farther the L\'evy flight tail index deviates from the Brownian limit. Our counterexample is based on the framework of the classic narrow capture problem in which a random search is performed for a small target within a confined search domain. Our results are obtained via three avenues: Monte Carlo simulations of the discrete search processes, finite difference solutions and a matched asymptotic analysis of the elliptic (pseudo)-differential equations of the corresponding continuum limits. Asymptotic analysis of the L\'evy search yields an expression for the average search time accurate to $\mathcal{O}(1)$, providing insights into how the latter is impacted by various features of the target and search domain.
\end{abstract}

\maketitle


\section{\label{sec:intro}Introduction}

It is a widely held belief that random search algorithms using L\'evy flights can find a target faster than using Brownian motion \cite{shlesinger1986levy, viswanathan1999optimizing, viswanathan2008levy, sims2019optimal}, \blue{with \cite{sims2019optimal} finding evidence for the intrinsic (i.e., arising from endogenous neurophysiological processes \cite{sims2019optimal, kolzsch2015experimental, sims2014hierarchical, de2011levy, humphries2012foraging,wearmouth2014scaling,maye2007order}) generation of L\'evy-like movement patterns in certain organisms. In \cite{gutierrez2015neural}, a neural coding scheme is used to explore a possible explanation for such search trajectories.} 

This so called ``L\'evy flight foraging hypothesis'' forms the basis of many biological models (e.g., \cite{viswanathan1996levy, reynolds2009honeybees}) as well as numerical search algorithms \cite{viswanathan1999optimizing, kaidi2022dynamic, yang2010eagle, yang2009cuckoo, heidari2017efficient}. It has also led the seeking of optimal L\'evy tail indices (e.g., \cite{viswanathan1999optimizing,levernier2020inverse}) for maximizing the amount of sparsely spaced targets being captured relative to the distance traversed \cite{viswanathan1999optimizing}, resulting in the 2020-21 dialogue that took place on Phys. Rev. Lett. \cite{levernier2020inverse, buldyrev2021comment, levernier2021reply}. 

In contrast to these existing works, we present an alternative means of quantifying efficiency via measuring the expected search time of a small, \blue{stationary} target in a finite domain and provide an example in two-dimensions for which the Brownian search strategy is more efficient than strategies based a L\'evy flight of any tail index $\alpha$. In fact, we demonstrate that in our setting, a certain power-law dependence of the expected search time on $\alpha$ that worsens the farther $\alpha$ deviates from its Brownian limit. 

The framework we employ for this comparison is consistent with that of first passage time problems for Brownian search (e.g., \cite{redner2001guide, benichou2015mean,benichou2014first,tejedor2009global}) in a geometry motivated by the narrow capture problem used to model biological and ecological processes (e.g., \cite{schuss2007narrow, holcman2015stochastic, holcman2014narrow, singer2006narrow, bressloff2021asymptotic, pillay2010asymptotic, kurella2015asymptotic, delgado2015conditional}). \blue{We emphasize here that, as is typical of narrow capture (and narrow escape) problems, we assume the search target is small and stationary. From a modeling perspective, the latter assumption of stationary targets means that, for example, in the modeling of animal foraging, we must assume that the search objective either does not move or moves very slowly in comparison to the searcher. The former scenario may arise when animals forage for patches of food (such as grass) or water sources. The latter scenario may be pertinent if the foraging predator diffuses on a much faster time scale than does the prey (see, e.g., \cite{kurella2015asymptotic}) so that the target may be approximated as stationary. Brownian search times of mobile targets are considered in, e.g., \cite{tzou2014first, tzou2015mean, lindsay2017optimization, gabel2012can, lawley2019diffusive}.}

We develop three different approaches to arrive at our result. First, we devise and implement a Monte Carlo simulation to calculate the expected search time of searches based on L\'evy flight. Second, we implement a numerical method for solving (pseudo)-differential equations which yields detailed information about the expected search time as a function of initial position. We use this numerical solution to gain insight into the potential mechanism behind why Brownian searches appear to take less time than L\'evy flights. Third, in \S \ref{appB}, we develop a matched asymptotic analysis to derive leading order analytic predictions for the expected search times. We remark that the comprehensive set of results from these three different approaches provide more quantitative and qualitative insight than the analytic asymptotic estimate for the search time discussed recently in \cite{chaubet2022geodesic}.

\begin{figure}[!htp]
	\centering
	\includegraphics[width=0.4\textwidth]{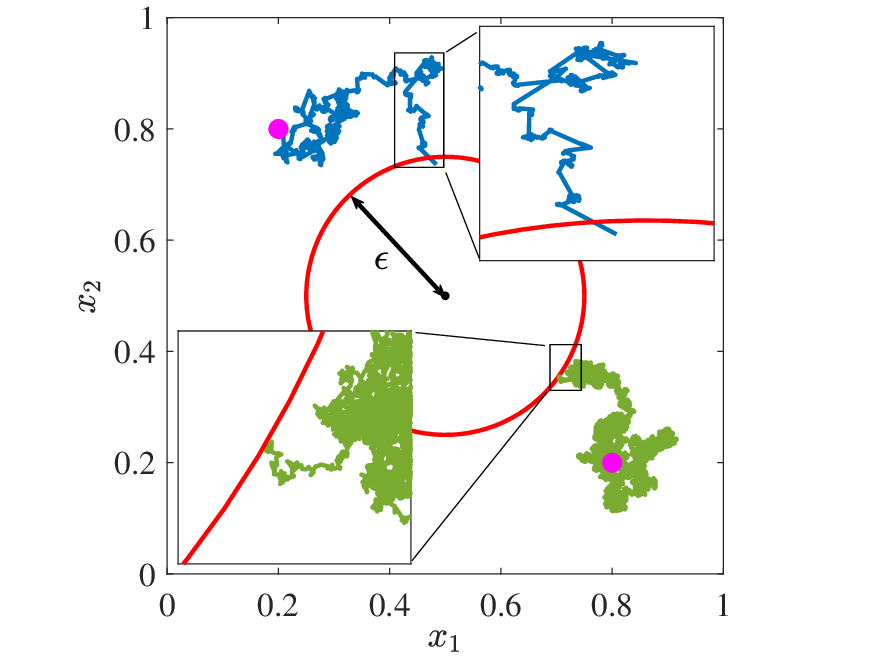}
	\caption{Illustration of L\'evy flight (blue) and Brownian walk (green) searches for a target of disk of radius $\eps = 0.25$ (red circle indicates its boundary) centered at the point $(0.5, 0.5)$ on the flat torus $\mathbb{T}^2$. The pink markers indicate the starting locations of each search. Each search continues until it first reaches the target. Closeups of both paths are shown in insets. The L\'evy search path (tail index $\alpha = 0.5$) exhibits series of small jumps separated by occasional long jumps while the Brownian path lacks the latter. Note that $\eps$ is set relatively large for illustrative purposes; we consider a smaller range of $\eps$ in this article.} 
	\label{random_walk_comparison}
\end{figure}

A schematic of the narrow capture framework for the geometry we consider is shown in Fig. \ref{random_walk_comparison}. The search domain is the unit torus $\mathbb{T}^2$ of unit side length with periodic boundary conditions and bottom left vertex at the origin. Two instances are shown of paths traced out by a Brownian (green) and L\'evy (blue) search for a \blue{stationary} target of disk of radius $\eps$ centered at $\bx_0 = (1/2, 1/2)$ (red). The pink dots mark the locations $\bx$ from where the respective searches begin. The search ends when the search first lands either on the boundary or inside the target disk.

The qualitative differences in the two paths are due to the probability distributions of their respective jump lengths. For the Brownian search, jump lengths are normally distributed with zero mean and variance of $\Delta t$ sufficiently small, leading to the linear-in-time mean squared displacement $\langle |\Delta \bx|^2 \rangle \propto \Delta t$. In the L\'evy search with tail index $0<\alpha<1$, jump lengths $|\by|$ are given by $|\by| = (\Delta t)^{1/(2\alpha)}|\bk|$, where $\bk$ is distributed according to a power-law distribution with tail $\sim |\bk|^{-(2+2\alpha)}$ (see, e.g., \cite{metzler2000random, valdinoci2009long, dipierro2022efficiency, metzler2004restaurant, benichou2015mean} and references therein). This leads to an unbounded mean squared displacement, and the super-linear scaling  $\langle |\Delta \bx|^{2\delta} \rangle^{1/\delta} \propto (\Delta t)^{1/\alpha}$ for $\delta < \alpha$. 

In the next section, we present our main findings and give possible reasons for the inferiority of L\'evy search strategies within the narrow escape framework.

\section{Main results and interpretation}

For a random search on $\mathbb{T}^2$, let us denote $u_\eps(\bx)$ ($v_\eps(\bx)$) the average search time (i.e., mean first passage time, or MFPT) of a L\'evy (Brownian) search starting from location $\bx$, and $B_\eps(\bx_0)$ the circular target of radius $\eps$ centered at $\bx_0$. Then the \textit{global} mean first passage time (GMFPT) \cite{tejedor2009global} is the expected search time averaged uniformly over all starting points $\bx \in \mathbb{T}^2 \setminus B_\eps(\bx_0)$. That is, the GMFTP, $\bar{u}_\eps$, of the L\'evy search is given by $\bar{u}_\eps = \int_{\mathbb{T}^2 \setminus B_\eps(\bx_0)} \! u_\eps(\bx)\,d\bx$, and similarly for the GMFPT, $\bar{v}_\eps$, of the Brownian search. 

\begin{figure}[!ht]
	\centering
	\includegraphics[width=0.36\textwidth]{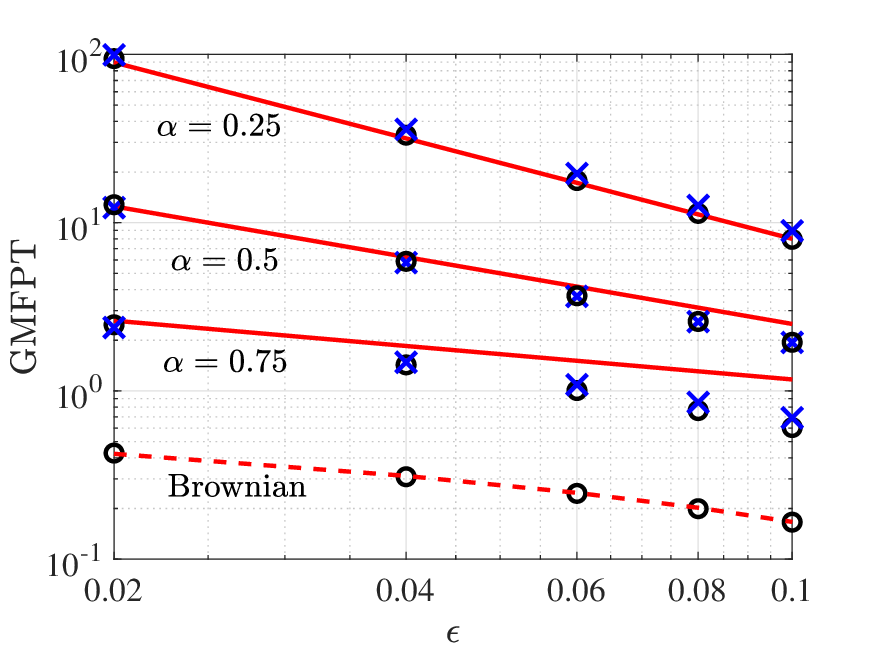}
	\caption{ (a) For L\'evy flight tail index $\alpha = 0.25$, $\alpha = 0.5$, $\alpha = 0.75$, and Brownian motion, with target of radius $\eps$ centered at $(1/2,1/2)$, we plot the GMFPT versus $\eps$. The results indicate that Brownian search is faster on average than the three L\'evy searches considered, and that average search times increase further the tail index $\alpha$ is from its Brownian limit of $1$. The quantities plotted in blue \texttt{x}'s and black \texttt{o}'s are obtained from Monte Carlo simulations and numerical solutions of the corresponding elliptic (pseudo)-differential equations, respectively. The solid red lines indicate the leading order power-law scaling of \eqref{ubarasympt}, while the dashed red curve plots the functional form $-(2\pi)^{-1}\log(\eps) + c$ for some constant $c$.}
	\label{mfpt_vs_eps_scale_final_all_three}
\end{figure}

We show in Fig. \ref{mfpt_vs_eps_scale_final_all_three} our primary result demonstrating that Brownian search is on average faster than L\'evy search. Moreover, the L\'evy flight search time increases the more its tail index $\alpha$ deviates from its Brownian limit of $1$. In the figure, we plot the L\'evy search GMFPT $\bar{u}_\eps$ for three different tail indices and for a range of target sizes $\eps$ on a log-log scale. The blue \texttt{x}'s are obtained from Monte Carlo simulations (see Fig. \ref{random_walk_comparison}), which we discuss below. The black \texttt{o}'s are obtained from finite difference solutions of the elliptic pseudo-differential equation for $u_\eps(\bx)$ given by \eqref{integral equation} corresponding to the continuum limit of the Monte Carlo process. Both confirm the leading order analytic result (shown in red) derived via a matched asymptotic analysis of \S \ref{appB}, stating that in the limit $\eps \to 0^+$ with $0<\alpha<1$,
\begin{equation}\label{ubarasympt}
	\begin{aligned}
		\bar{u}_\eps &\sim  A_\alpha\eps^{2\alpha -2} - R_\alpha(\bx_0;\bx_0) + \mo(1)  \,; \\ \quad A_\alpha &:= \frac{\Gamma(1-\alpha) (1-\alpha)}{4^\alpha \Gamma(\alpha) \sin((1-\alpha)\pi)} \,. 
	\end{aligned}
\end{equation}
In \eqref{ubarasympt}, $R_\alpha(\bx_0;\bx_0)$ is the regular part of a certain Green's function satisfying \eqref{G} evaluated at the location of the singularity, and is given by \eqref{Ralpha}. 

We make several remarks on the result \eqref{ubarasympt}. Firstly, $A_\alpha$ in \eqref{ubarasympt} is a function of the geometry of the target (in our instance, a disk of radius $\eps$) and the size of the search domain (in our instance, unity). Both of these dependencies are made explicitly clear in our derivation of \eqref{ubarasympt} in \S \ref{appB}. Secondly, the fact that target geometry effects enter at the leading order of $\bar{u}$ is in direct contrast to the Brownian motion result, where target geometry effects enter at $\mathcal{O}(1)$, which is the second order. Lastly, the leading order term in \eqref{ubarasympt} contains no information on global geometric properties of the search domain other than its size. These global effects are encoded only in the $\mathcal{O}(1)$ correction term, $R_\alpha(\bx_0;\bx_0)$.

The error of the approximation \eqref{ubarasympt} grows as $\alpha$ nears its Brownian limit of $1$, accounting for the worsening discrepancies between the blue \texttt{x}/black \texttt{o} and the (red) analytic prediction as $\alpha$ gets closer to $1$. We also plot $\bar{v}_\eps$, the GMFPT of the Brownian search, obtained from numerically solving \eqref{veps} (black \texttt{o}'s). The red dashed curve plots the functional form $-(2\pi)^{-1}\log(\eps) + c$ for some constant $c$, confirming the well-known the leading order $\mathcal{O}(|\log\eps|)$ scaling of Brownian search times (see, e.g., \cite{schuss2007narrow, holcman2015stochastic, holcman2014narrow, singer2006narrow, bressloff2021asymptotic, pillay2010asymptotic}).

\begin{figure}[!htbp]
	\centering
	\begin{subfigure}[b]{0.36\textwidth}
		\centering
		\includegraphics[width=\textwidth]{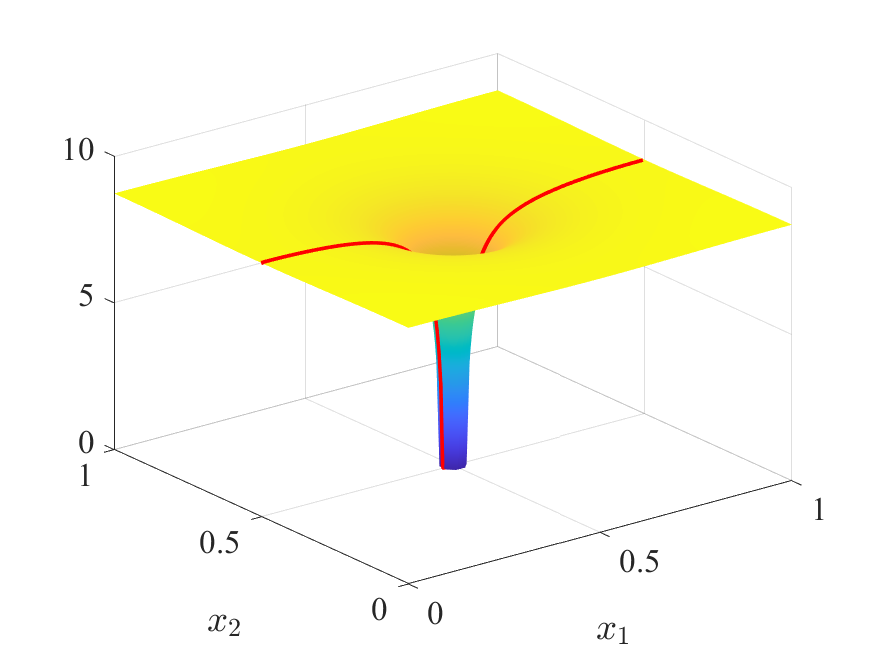}
		\caption{$u_\eps(p)$}
		\label{eps_0p03_surf_full}
	\end{subfigure}
	\begin{subfigure}[b]{0.36\textwidth}
		\centering
		\includegraphics[width=\textwidth]{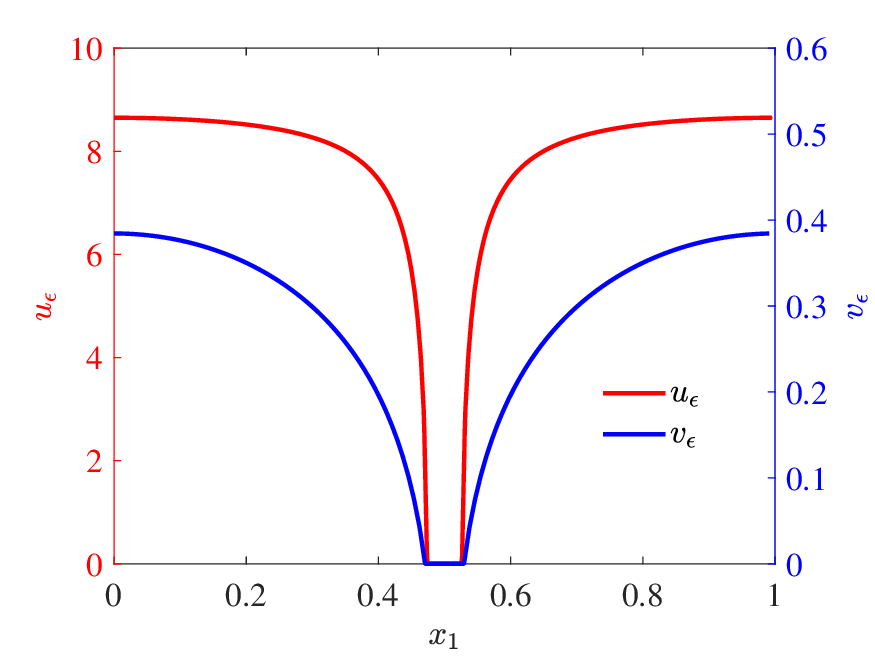}
		\caption{horizontal cross-section of $u_\eps$, $v_\eps$}
		\label{eps_0p03_line_full}
	\end{subfigure}
	\caption{For $\alpha = 1/2$ with target of radius $\eps = 0.03$ centered at $(1/2,1/2)$, we show in (a) the numerical solution for $u_\eps$ of \eqref{integral equation}. The red line indicates the contour plotted in red and (b), where, in blue, we also plot the corresponding contour of the numerical solution $v_\eps$ of \eqref{veps}. Note that $u_\eps$ ($v_\eps$) is plotted on the left (right) vertical axis. The spatial average of $u_\eps$ is approximately $8.57$ while that of $v_\eps$ is approximately $0.357$, indicating that an average search conducted via the L\'evy process with $\alpha = 1/2$ for a small target will be significantly longer in comparison.}
	\label{eps_0p03}
\end{figure}

We give a possible explanation for the longer average duration of L\'evy searches. In Figs. \ref{eps_0p03}, we plot the finite difference solution for $u_\eps(\bx)$ when $\alpha = 1/2$ and $\eps = 0.03$, and compare a cross section of this solution to that of $v_\eps(\bx)$. In Fig. \ref{eps_0p03_line_full}, we first observe that both solutions are identically $0$ for $\bx \in B_\eps(\bx_0)$ (i.e., searches beginning in and on the target cost zero time). Near the target boundary, we observe a much sharper rise in $u_\eps$ than for $v_\eps$, while far from the target, $u_\eps$ is flatter than $v_\eps$. This behavior of $u_\eps$ suggests that proximity to the target of starting location has little impact on the L\'evy search time. This owes to there being a greater likelihood of taking a long jump in the ``wrong'' direction, especially when the target is small.

\begin{figure}[!htp]
	\centering
	\begin{subfigure}[b]{0.36\textwidth}
		\centering
		\includegraphics[width=\textwidth]{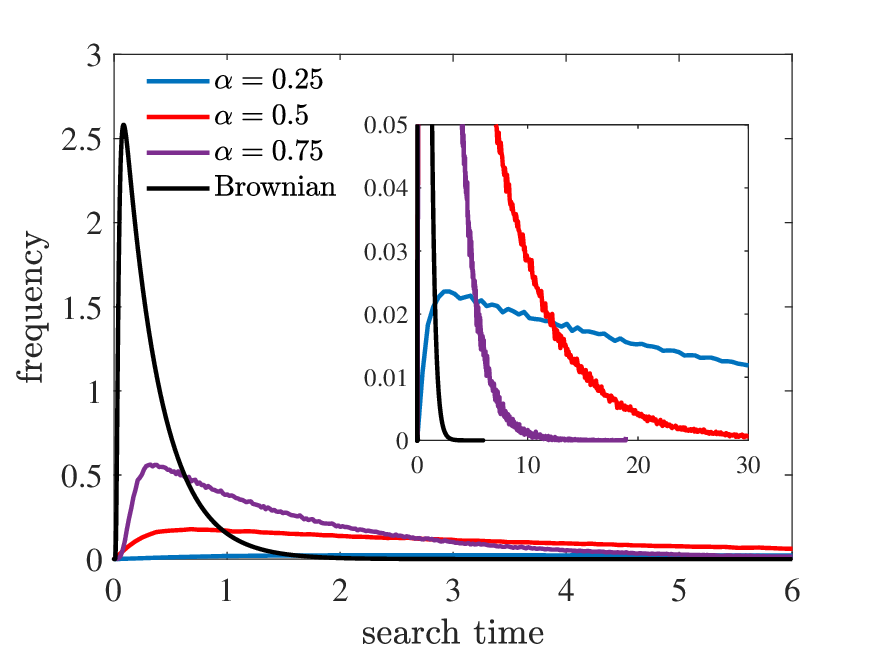}
		\caption{distribution of search times}
		\label{hist_alpha_0p25_0p5_0p75_brown}
	\end{subfigure}
	\begin{subfigure}[b]{0.36\textwidth}
		\centering
		\includegraphics[width=\textwidth]{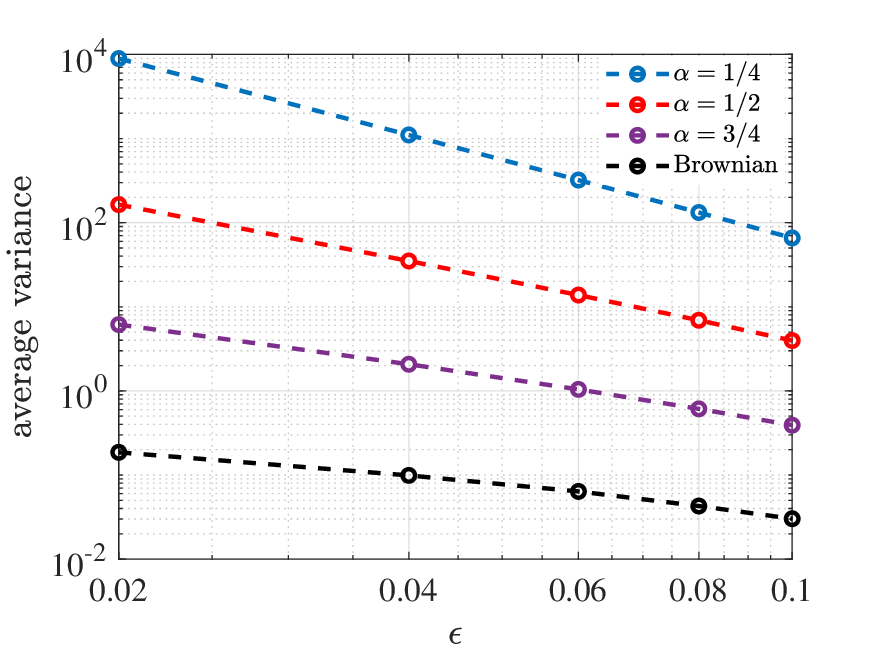}
		\caption{average variance}
		\label{variance_fig_all}
	\end{subfigure}
	\caption{(a) Starting from the point $(0,0)$, the figure shows the full probability density function of search times for a target of radius $\eps = 0.04$ centered at $p_0 = (1/2, 1/2)$. Results for the Brownian search are obtained from numerically solving the time-dependent diffusion PDE while those for the L\'evy searches are obtained from Monte Carlo simulations. Observe the increasingly heavy tail of the distributions as $\alpha$ decreases. This indicates the increasing likelihood of long search times, contributing to the longer average search times of L\'evy flights relative to Brownian walks. (b) Finite difference computation of search time variance averaged over all starting locations, obtained from the second moments of MFPT in \eqref{sigmaeps} and \eqref{taueps}. } 
\end{figure}

Thus, a L\'evy search that has reached the vicinity of the target may take a long jump away from it, effectively forcing it to restart its search from a farther location. Repeated approaches to the target followed by long jumps away from it can lead to anomalously long search times, which we show in Fig. \ref{hist_alpha_0p25_0p5_0p75_brown}.	Obtained from Monte Carlo simulations of search processes beginning at $(0,0)$, the probability density distributions of L\'evy search times differ greatly from that of Brownian search times. In particular, the comparatively slow decay of the tail for longer search times is apparent, especially for $\alpha = 1/4$. The greater variance of L\'evy search times is confirmed in Fig. \ref{variance_fig_all}, where we show finite difference computations of the search time variance averaged over all starting locations. We note the near-linear behavior of L\'evy flight variances, suggesting that they, along with the GMFPT $\bar{u}_\eps$, follow a power-law scaling.

In the following sections, we provide an overview of our methodology, beginning with a brief outline of our Monte Carlo algorithm . We then give the elliptic pseudo-differential equations which characterize the continuum limit of the L\'evy search process. Mathematical details are presented in \S \ref{methods}. We close by discussing some open problems.

\section{Monte Carlo Simulation of L\'evy Flight Search on $\mathbb T^2$}\label{monte carlo}

We describe here the Monte Carlo algorithm that we used to generate the simulated L\'evy flight search times of the previous section. \blue{The algorithm is motivated by the description of L\'evy flights on $\mathbb R^n$ given by Valdinoci \cite{valdinoci2009long}. We remark that the purpose of \cite{valdinoci2009long} is to describe the derivation of the (continuous) fractional heat equation from a discrete L\'evy flight framework. It is this discrete framework that we invoke in \S \ref{appA} to derive the pseudo-differential equation that we subject to analysis in \S \ref{appB}, resulting in the main result \eqref{ubarasympt}. We base the simulations on this discrete framework. Other algorithms for simulating stable L\'evy process are given in, e.g., \cite{mantegna1994fast, weron2005computer, chambers1976method}.}

We describe the process on $\mathbb R^2$ from which we can derive the process on $\mathbb T^2$ simply by identifying $(n+r, m+s) \equiv (r,s)$ for $n, m\in \mathbb Z$ and $r,s\in (0,1)$. For the $(j+1)-th$ step starting from $\bx_j \in \mathbb{T}^2$ with $j = 0,1, 2, \ldots$, the $j$-th displacement of the L\'evy flight with tail index $0<\alpha < 1$ is given by $h\bk$, where $h$ is a sufficiently small parameter (in particular, sufficiently smaller than the size of the target) and $\bk \in \mathbb{Z}^2$ is a random variable drawn from a power-law distribution with tail $\sim~|\bk|^{-2-2\alpha}$ (see \S \ref{rejection}). Since $\langle |\bk|^{2\delta} \rangle$ is bounded for $0 <\delta < \alpha < 1$, we have the scaling of the expected displacement $\langle |\Delta \bx|^{2\delta} \rangle \propto h^\delta$, or $\langle |\Delta \bx|^{2\delta} \rangle^{1/\delta} \propto h$. We then let $\Delta t = D_\alpha h^{2\alpha}$ so that  $\langle |\Delta \bx|^{2\delta} \rangle^{2/\delta} \propto (\Delta t)^{1/\alpha}$ as required. Here, $D_\alpha$ is a constant chosen so that the process is consistent with Brownian motion in the limit $\alpha \to 1^-$ (see \eqref{Dalpha} of \S \ref{appA}). As mentioned above, from this jump process on $\mathbb R^2$ one can deduce the process on $\mathbb T^n$ simply by modding out $\mathbb Z^2$.

In \S \ref{rejection}, we describe our process for sampling from a discrete power law distribution. We then describe, in \S \ref{monto_carlo_full}, the Monte carlo simulation of the search process from which we generate the Monte Carlo estimates for search times in Fig. \ref{mfpt_vs_eps_scale_final_all_three}.


\subsection{Rejection sampling algorithm for $\mathcal{K}_\alpha$} \label{rejection}

We describe here the rejection sampling algorithm that we use to draw  samples $\bk \in \mathbb{Z}^2$ from the discrete power law distribution
\begin{subequations} \label{kappa}
	\begin{equation} \label{K}
		\mathcal{K}_\alpha(\mathbf{k}) = \begin{cases} 
			0 & \mathbf{k} = \mathbf{0} \\
			C_\alpha|\mathbf{k}|^{-2-2\alpha} &  \mathbf{k} \neq \mathbf{0}
		\end{cases} \,,
	\end{equation}
	where $C_\alpha$ is the normalizing constant given by
	\begin{equation} \label{Calpha}
		C_\alpha = \frac{1}{\sum\limits_{\mathbf{k} \in \mathbb{Z}^2 \,, \, \mathbf{k} \neq \mathbf{0}} |\mathbf{k}|^{-2-2\alpha} } \,,
	\end{equation}
\end{subequations}
where $\bk \in \mathbb{Z}^2$.

First observe that 
\begin{eqnarray}
	\label{majorizing}\mathcal{K}_\alpha(\bk) \leq \frac{C_\alpha}{\tilde C_\alpha} \tilde{\mathcal{K}_\alpha}(\bk) \,,
\end{eqnarray}
where $\tilde{\mathcal{K}}_\alpha (\bk) =  \tilde C_\alpha|\bk|_\infty^{-2-2\alpha}$ for $\bk \in \mathbb{Z}^2 \setminus \lbrace \mathbf{0} \rbrace$, $\tilde{C}_\alpha$ is the normalization constant (given below), and $|\cdot|_\infty$ is the $\ell^\infty$ norm on $\mathbb R^2$. The distribution $\tilde {\mathcal{K}_\alpha}$ serves as a good proposal distribution for rejection sampling because the process of sampling from it is relatively simple, which we describe now.


The distribution $\tilde{\mathcal{K}}_\alpha(k)$ depends purely on the $\ell^\infty$ norm of the random variable $\bk\in \mathbb Z^2$. As such we observe for each fixed $\hat \bk\in \mathbb Z^2$, a random variable $\bk\in \mathbb Z^2$ satisfies
$$\mathbb P(\bk = \hat \bk) = \mathbb P(|\bk|_\infty = |\hat \bk|_\infty) \mathbb P(\bk = \hat \bk \mid |\bk|_\infty = |\hat \bk|_\infty) \,,$$
where $\mathbb P(\bk = \hat \bk \mid |\bk|_\infty = |\hat \bk|_\infty)$ is uniformly distributed amongst the $8|\hat \bk|_\infty$ points having $\ell^\infty$ norm $|\hat \bk|_\infty$. 

For each $n\in \mathbb N$, using the explicit form of $\tilde{\mathcal{K}}_\alpha$ and the fact there are $8n$ points on $\mathbb Z^2$ having $\ell^\infty$ norm $n$, we see that
\begin{eqnarray}
	\label{probability of norm}
	\mathbb P(|\bk| = n) =\tilde C_\alpha 8n/n^{2+2\alpha} =\tilde C_\alpha 8/n^{1+2\alpha} \,,
\end{eqnarray}
where $\tilde{C}_\alpha$ is given by 
\begin{equation} \label{Catilde}
	\tilde C_\alpha = \frac{1}{\sum\limits_{n=1}^\infty  8n^{-1-2\alpha}} \,.
\end{equation}

\noindent This distribution can be sampled from using inversion sampling for discrete distributions.

\begin{rmk}
	We observe that in the special case when $\alpha = 1/2$ we can derive explicit analytic expressions for $\mathbb P(|k| = n)$. Indeed, we can sum over $n$ in \eqref{probability of norm} to yield the condition
	
	$$ 1 =  8\tilde C_{1/2}\sum_{n=1}^\infty \frac{1}{n^{2}} = \frac{4\pi^2}{3} \tilde C_{1/2}.$$
	That is, $\tilde C_{1/2}= \frac{3}{4\pi^2}$. Inserting this back to \eqref{probability of norm} we get, for each fixed $n\in \mathbb N$,
	\begin{eqnarray}
		\label{sampling the radius}
		\mathbb P(|\bk| = n) = \frac{6}{\pi^2} \frac{1}{n^{2}} \,,
	\end{eqnarray}
	for $\alpha = 1/2$. 
\end{rmk}

The rejection sampling algorithm for the discrete power law distribution $\mathcal{K}_\alpha$ is thus as follows:

\begin{enumerate}
	\item sample $n\in \{1, \dots,  10000\}$ from \eqref{probability of norm} using inversion sampling;
	\item for this $n\in \mathbb N$ sample $\bk\in \mathbb Z^2$ uniformly from the $8n$ points on $\mathbb Z^2$ have $\ell^\infty$ norm $n$;
	\item for this $\bk\in \mathbb Z^2$, sample $r\in \left(0, \frac{C_{\alpha}}{\tilde C_{\alpha}} \tilde{\mathcal{K}_\alpha}(\bk)\right )$ uniformly. If $r \leq \mathcal{K}_\alpha(\bk)$, accept this $\bk\in \mathbb Z^2$. If not, reject and repeat.
\end{enumerate}
Numerical experiments show that this rejection sampling algorithm accepts approximately $69\%$ of the time.

\subsection{Monte Carlo algorithm} \label{monto_carlo_full}

We describe here the Monte Carlo algorithm used to simulate L\'evy flight searches.Let $\mathbb{T}^2$ be the flat torus $[0,1]\times[0,1]$ with periodic boundary conditions, and $u_\epsilon^{disc}(\bx)$ be the expected discrete L\'evy flight search time (obtained via simulation of a discrete process) of a circular target $B_\epsilon(\mathbf{0})$ of radius $\epsilon > 0$ centered at $(1/2, 1/2)$ starting from $\bx \in \mathbb{T}^2$. Then for a L\'evy flight tail index $\alpha \in (0,1)$ and $h > 0$ sufficiently small, we perform the following Monte Carlo procedure to compute an approximation of $u_\epsilon^{disc}(\bx)$:\newline\newline
\noindent Set $T = 0$ and $\bx = (x_1, x_2)$.\newline\newline
\noindent Repeat the following until $\bx\in B_\epsilon(\mathbf{0})$:
\begin{enumerate}
	\item sample $\bk\in \mathbb Z^2$ from $\mathcal{K}_\alpha$   using the above rejection sampling algorithm;
	\item set $\by := (y_1, y_2) = \bx + h\bk$;
	\item set $\bx = (y_1 \bmod \lfloor y_1 \rfloor, y_2 \bmod \lfloor y_2 \rfloor)$, which accounts for the periodic boundary conditions of $\mathbb{T}^2$;
	\item set $T = T + \Delta t$, where $\Delta t  = D_\alpha h^{2\alpha}$, with $D_\alpha$ is given in \eqref{Dalpha} of \S \ref{appA}.
\end{enumerate}
The $j$-th run of the above generates a stopping time $T_j$. After executing $M$ runs and generating stopping times $T_1, T_2, \dots, T_M$ we calculate 
\begin{eqnarray}
	\label{E}
	u_\epsilon^{disc}(\bx)  \approx \frac{T_1 + T_2 + \dots + T_M}{K} \,,
\end{eqnarray}
for large $M\in \mathbb N$. Repeating this process over a grid of points $\bx \in \mathbb{T}^2$ and then averaging, we are able to obtain an approximation of the global mean first passage time (GMFPT); i.e., the spatial average of $u_\epsilon^{disc}(\bx)$ over $\mathbb{T}^2$.

\section{The Elliptic (Pseudo)-Differential Equations}\label{numerics}

In this section, we briefly discuss the elliptic pseudo-differential equation satisfied by $u_\eps(\bx)$, the MFPT of the L\'evy search starting from point $\bx \in \mathbb{T}^2$ in the continuous limit. Adopting the electrostatics approach of \cite{redner2001guide} for Brownian searches, we show in \S \ref{appA} that $u_\eps(\bx)$ satisfies the \textit{exterior problem}
\begin{subequations} \label{Au} \begin{equation} \label{integral equation}
		\mathcal{A}_\alpha u_\eps = -1\ {\rm on}\ \mathbb T^2\setminus \overline {B_\eps(\bx_0)},\ \ u_\eps = 0\ {\rm on }\ \overline{B_\eps(\bx_0)} \,,
	\end{equation}
	where $\mathcal{A}_\alpha$ for $0<\alpha<1$ is the fractional Laplacian of order $\alpha$ on $\mathbb{T}^2$ given by \begin{multline}  \label{fraclap}
		\mathcal{A}_\alpha f(\bx) :=  \frac{4^\alpha \Gamma(1+\alpha)}{\pi |\Gamma(-\alpha)|}\times \\ \int\limits_{[0,1]\times [0,1]} \! \sum_{\mathbf{m} \in \mathbb{Z}^2} \frac{f(\by) - f(\bx)}{|\bx-(\by + \mathbf{m})|^{2+2\alpha}}  \, d\by\,,
	\end{multline}
where we have identified $\mathbb T^2$ isomorphically with $[0,1)\times[0,1)$ so that the expression $|\bx-(\by + \mathbf{m})|$ is well-defined.	
	
We remark that the eigenvalues of $\mathcal{A}_\alpha$ converge to that of the usual (local) Laplacian on $\mathbb{T}^2$ as $\alpha \to 1^-$. Furthermore, for a L\'evy flight on $\mathbb{R}^2$, the operator $\mathcal{A}_\alpha$ would simply become the standard fractional Laplacian of order $\alpha$ given in \eqref{fraclap} of \S \ref{appA}. The lattice sum kernel of \eqref{integral equation} sums the probabilities of all the possible paths from $\by$ to $\bx$ on $\mathbb{T}^2$. By analogy with \cite{redner2001guide, kurella2015asymptotic}, we have that the second moment of the MFPT (SMFPT), $w_\eps(\bx)$, satisfies \begin{equation}\label{sigmaeps} \mathcal{A}_\alpha  w_\eps = -2 u_\eps(\bx) \ {\rm on}\ \mathbb T^2\setminus \overline {B_\eps(\bx_0)},\ \ w_\eps = 0\ {\rm on }\ \overline{B_\eps(\bx_0)} \,.\end{equation}\end{subequations} The boundary value problems for the Brownian search time ($v_\eps(\bx)$) and SMFPT ($\tau_\eps(\bx)$)  are well-known (\cite{redner2001guide}): \begin{subequations} \begin{equation} \label{veps} \Delta v_\eps = -1\ {\rm on}\ \mathbb T^2\setminus \overline {B_\eps(\bx_0)},\ \ v_\eps = 0\ {\rm on }\ \partial B_\eps(\bx_0) \,,\end{equation}\begin{equation} \label{taueps} \Delta \tau_\eps = -2v_\eps\ {\rm on}\ \mathbb T^2\setminus \overline {B_\eps(\bx_0)},\ \ \tau_\eps = 0\ {\rm on }\ \partial B_\eps(\bx_0) \,.\end{equation} \end{subequations} 
The \textit{variance} of the MFPT (VMFPT) is then given by $\textrm{VMFPT} = \textrm{SMFPT} - \textrm{MFPT}^2$, plotted in Fig. \ref{variance_fig_all}. Finite difference solutions of \eqref{integral equation}-\eqref{taueps} are straightforward, although discretizing the operator $\mathcal{A}_\alpha$ is computationally expensive. As such, asymptotic methods for equations of the forms \eqref{integral equation} and \eqref{sigmaeps}, such as that provided in \S \ref{appB}, may be helpful in reducing computation requirements.

\section{Asymptotic solution of $u_\eps(\bx)$} \label{methods}

In \S \ref{appA}, we provide a derivation of the pseudo-differential equation \eqref{integral equation} for $u_\eps(\bx)$ starting from a random walk formalism. In \S \ref{appB}, we provide an two-term asymptotic solution for $u_\eps$ leading to the main result \eqref{ubarasympt} using a matched asymptotic analysis. Through this analysis, we highlight how effects of target and search domain geometry enter into the expression for the GMFPT. In \S \ref{appC}, we discuss how this analysis extends to the flat 3-D torus as well as closed 2-D manifolds. 

\subsection{Derivation of elliptic (pseudo)-differential equations for average Brownian and L\'evy search times} \label{appA}

We begin with the derivation of \eqref{Au}, the pseudo-differential equation for the average search time via L\'evy flight of a circular target of radius $\eps$ starting from some point $\bx = (x_1,x_2)$. The derivation is based on a continuum limit of a discrete process, and is given in \cite{redner2001guide} for a Brownian search, which we reproduce first for the purpose of completeness.

Consider a Brownian particle on an $N\times N$ grid with spacing $h = 1/(N-1)$. At regular intervals of $\Delta t$, the particle hops from its current location $(x_1,x_2)$ to one of its four neighboring points, $(x_1+h,x_2)$, $(x_1-h,x_2)$, $(x_1,x_2+h)$, and $(x_1,x_2-h)$ with equal probability. Here the addition by $\pm h$ are understood to be carried out mod $1$. Let $v_\eps(\bx)$ be the expected search time starting from $(x_1,x_2)$. Then $v_\eps(\bx)$ must  be the average of the expected search time starting from one of its four neighboring points, plus the $\Delta t$ traversal time. That is,
\begin{multline} \label{brown1}
	v_\eps(x_1,x_2) = \frac{1}{4}\left[
		v_\eps(x_1+h,x_2) + v_\eps(x_1-h,x_2) + \right. \\ \left. v_\eps(x_1,x_2+h)  + v_\eps(x_1,x_2-h) \right] + \Delta t \,.
\end{multline}
Dividing \eqref{brown1} by $\Delta t$ and rearranging, we have
\begin{multline}\label{brown2}
	\frac{1}{4\Delta t}\left\lbrack v_\eps(x_1+h,x_2) + v_\eps(x_1-h,x_2)  + v_\eps(x_1,x_2+h) \right. \\ \left. + v_\eps(x_1,x_2-h) - 4v_\eps(x_1,x_2) \right\rbrack + 1 = 0\,.
\end{multline}
Taking the limit $h \to 0$ and $\Delta t \to 0$  in \eqref{brown2} while maintaining $h^2/(4\Delta t) = 1$, we obtain 
\begin{equation}\label{brown3}
	\Delta v_\eps + 1 = 0\,, \quad \bx \in \mathbb{T}^2 \,; \qquad v_\eps = 0 \,, \quad \bx \in \partial B_\eps(\bx_0) \,,
\end{equation}
where the boundary condition of \eqref{brown3} is the statement that the search time starting from the boundary of the target is zero. For the unit torus $\mathbb{T}^2$, $v_\eps(\bx)$ would satisfy periodic conditions on the boundary.

Following the idea of \cite{valdinoci2009long}, we now generalize this derivation to L\'evy flights characterized by a jump length distribution whose tail decays according to the power law $\sim |\mathbf{k}|^{-2-2\alpha}$ for $0 < \alpha < 1$. For the flat torus $\mathbb{T}^2$ with $\bx \in [0,1)\times [0, 1)$, we have the same $N\times N$ grid with spacing $h = 1/(N-1)$.  Instead of \eqref{brown1} which allows only for nearest-neighbor jumps, we have
\begin{equation} \label{levy1}
	u_\eps(h\bi)= \sum_{j_1 = 0}^{N-1}\sum_{j_2 = 0}^{N-1} w(h\bi,h\bj)u_\eps(h\bj) + \Delta t \,.
\end{equation}
In \eqref{levy1}, $\bi = (i_1, i_2) \in [0,N-1]\times[0,N-1]$, $\bj = (j_1, j_2)$, and $w(h\bi, h\bj)$ is a sum of the probabilities of all possible paths  from $h\bi$ to $h\bj$ on $\mathbb{T}^2$, taking into account that there are an infinite number of ways to travel from one point to another via a straight line owing to the periodic nature of $\mathbb T^2$. From \cite{valdinoci2009long}, a jump of $h\bk$ has the probability given by $\mathcal{K}_\alpha(\bk)$, where the discrete probability mass function $\mathcal{K}_\alpha(\mathbf{k})$ on $\mathbb{Z}^2$ is given by \eqref{kappa} of \S \ref{rejection}.

The probability of reaching $h\bi$ from $h\bj$ on $\mathbb{T}^2$ must then be given by the lattice sum
\begin{equation} \label{latticesum}
	w(h\bi, h\bj) = \sum_{\mathbf{m}\in \mathbb{Z}^2} C_\alpha|\bi - (\bj + \mathbf{m}(N-1)) |^{-2-2\alpha} \,,
\end{equation}
where $\mathbf{m} = (m_1, m_2) \in \mathbb{Z}^2$, and we have tiled $\mathbb{R}^2$ with the unit square, and $\bj + \mathbf{m}(N-1)$ is the point corresponding to $\bj \in \mathbb{T}^2$ in the square with bottom left vertex at $\mathbf{m}(N-1)$. In \eqref{latticesum}, $C_\alpha$ is the normalization constant of \eqref{Calpha}. Since \begin{equation}
	\sum_{j_1 = 0}^{N-1}\sum_{j_2 = 0}^{N-1} w(h\bi, h\bj) = \sum_{\bk \in \mathbb{Z}^2}C_\alpha|\bi - \bk|^{-2-2\alpha}  = 1  \,,
\end{equation}
\eqref{levy1} can be rewritten \begin{equation} \label{levy2}
	\sum_{j_1 = 0}^{N-1}\sum_{j_2 = 0}^{N-1} w(h\bi, h\bj)(\mathbf{k})\left[u_\eps(h\bj) - u_\eps(h\bi)\right] + \Delta t = 0 \,.
\end{equation}
Using the formal scaling law of \cite{valdinoci2009long},
\begin{equation} \label{scalinglaw}
	\Delta t = D_\alpha h^{2\alpha} \,,
\end{equation}
for some constant $D_\alpha$ to be determined, we divide both sides of \eqref{levy2} by $\Delta t$ to obtain
\begin{equation} \label{levy3}
	\frac{1}{D_\alpha}h^2\sum_{j_1 = 0}^{N-1}\sum_{j_2 = 0}^{N-1} w(h\bi,h\bj)\left[\frac{u_\eps(h\bj) - u_\eps({h\bi})}{h^{2\alpha+2}}\right] + 1 = 0 \,.
\end{equation}
Using \eqref{latticesum} in \eqref{levy3}, we obtain 
\begin{widetext}
	\begin{equation}\label{levy3b}
	\frac{C_\alpha}{D_\alpha}h^2\sum_{j_1 = 0}^{N-1}\sum_{j_2 = 0}^{N-1} \sum_{\mathbf{m}\in \mathbb{Z}^2} |\bi - (\bj + \mathbf{m}(N-1)) |^{-2-2\alpha}\left[\frac{u_\eps(h\bj) - u_\eps(h\bi)}{h^{2\alpha+2}}\right] + 1 = 0 \,.
\end{equation}
\end{widetext}
Letting $\bx \equiv h\bi$ and $\by \equiv h\bj$, and recalling that $h = 1/(N-1)$, \eqref{levy3b} becomes \begin{equation} \label{levy4}
	\frac{C_\alpha}{D_\alpha}h^2\sum_{j_1 = 0}^{N-1}\sum_{j_2 = 0}^{N-1} \sum_{\mathbf{m}\in \mathbb{Z}^2} \left[\frac{u_\eps(\by) - u_\eps(\bx)}{|\bx - (\by + \mathbf{m}) |^{2+2\alpha}}\right] + 1 = 0 \,.
\end{equation}
In the limit $h \to 0^+$, \eqref{levy4} is the Riemann sum approximation to the pseudo-differential equation
\begin{multline}\label{levy5}
	\frac{D_\alpha}{C_\alpha} \int\limits_{[0,1]\times [0,1]} \!  \left\lbrack u_\eps(\by) - u_\eps(\bx) \right\rbrack\\ \times \sum_{\mathbf{m} \in \mathbb{Z}^2} \frac{1}{|\bx-(\by + \mathbf{m})|^{2+2\alpha}}    \,d\mathbf{y} + 1 = 0 \,,
\end{multline}
which is the desired form given in \eqref{fraclap}. 

To motivate the selection of $D_\alpha$, we rewrite \eqref{levy5} in terms of the fractional Laplacian defined on $\mathbb{R}^2$ by considering the periodic extension of $u_\eps(\bx): \mathbb T^2 \to \mathbb R$, which we denote $U_\eps(\bx) : \mathbb R^2 \to \mathbb R$ defined by $U_\eps(x_1, x_2) = u_\eps((x_1 \bmod 1, x_2 \bmod 1))$. The equation \eqref{levy5} can now be written
\begin{equation} \label{levy6}
	\frac{D_\alpha}{C_\alpha} \int_{\mathbb{R}^2} \! \frac{U_\eps(\by) - U_\eps(\bx)}{|\bx-\by|^{2+2\alpha}}    \,d\mathbf{y} + 1 = 0 \,,
\end{equation}
It is shown in \cite{valdinoci2009long} that the integral term in \eqref{levy6} is well defined as a principal value integral when $\mathbf{x}$ is near $\mathbf{y}$:
\begin{equation} \label{pv}
	\int_{\mathbb{R}^2} \! \frac{U_\eps(\mathbf{y}) - U_\eps(\mathbf{x})}{|\bx-\mathbf{y}|^{2+2\alpha}} \,d\mathbf{y} = \lim_{\delta \to 0^+} \int\limits_{\mathbb{R}^2 \setminus B_\delta(\by)} \! \frac{U_\eps(\mathbf{y}) - U_\eps(\mathbf{x})}{|\bx-\mathbf{y}|^{2+2\alpha}} \,d\mathbf{y} \,,
\end{equation}
where $B_\delta(\by)$ is the ball of radius $\delta$ centered at $\by$. With the fractional Laplacian in $\mathbb{R}^2$ given by
\begin{equation} \label{fL}
	-(-\Delta)^\alpha f(\mathbf{x}) = \frac{4^\alpha \Gamma(1+\alpha)}{\pi|\Gamma(-\alpha)|}\int_{\mathbb{R}^2} \! \frac{f(\mathbf{y}) - f(\mathbf{x})}{|\mathbf{x}-\mathbf{y}|^{2+2\alpha}} \, d\mathbf{y} \,,
\end{equation}
\eqref{levy6} becomes
\begin{equation} \label{levy7}
	-\frac{D_\alpha}{C_\alpha} \frac{\pi|\Gamma(-\alpha)|}{4^\alpha \Gamma(1+\alpha)} (-\Delta)^\alpha U_\eps(\mathbf{x}) + 1 = 0 \,.
\end{equation}
Then setting 
\begin{equation} \label{Dalpha}
	D_\alpha = C_\alpha\frac{4^\alpha \Gamma(1+\alpha)}{\pi|\Gamma(-\alpha)|} \,,
\end{equation}
so that the coefficient in front of $(-\Delta)^\alpha U_\eps(\mathbf{x})$ is equal to one, we finally arrive at 
\begin{equation} \label{levy8}
	-(-\Delta)^\alpha U_\eps(\mathbf{x}) + 1 = 0 \,.
\end{equation}
Returning to $\mathbb{T}^2$, we recover \eqref{Au}.

In \eqref{Dalpha}, $C_\alpha$ is the normalization constant of \eqref{Calpha}. Note that the fractional Laplacian $-(-\Delta)^\alpha$ for $0<\alpha<1$ can be defined in terms of the Fourier transform $\mathcal{F}(\cdot)[\boldsymbol{\xi}]$ by $\mathcal{F}((-\Delta)^\alpha f(\mathbf{x}))[\boldsymbol{\xi}] = -|\xi|^{2\alpha} \mathcal{F}(f(\mathbf{x}))[\boldsymbol{\xi}]$. Thus, as $\alpha \to 1^-$, the Fourier multiplier of $-(-\Delta)^\alpha$ approach those of the usual Laplacian operator on $\mathbb R^n$

Note that the Fourier multiplier definition of the fractional Laplacian on $\mathbb R^n$ also allows us to deduce that the operator in \eqref{levy5} is also the fractional Laplacian on $\mathbb T^2$ as defined via a spectral decomposition. Indeed, it is well known that $e^{i\bk\cdot x}$ are eigenfunctions for the Laplacian on $\mathbb T^2$ for $\bk \in \mathbb Z^2$. To compute the action of the operator in \eqref{levy5} on $e^{i\bk\cdot x}$, we extend it periodically as smooth functions on $\mathbb R^2$ and apply the operator in \eqref{levy6} to the extension. Observing that it has a Fourier multiplier representation, we see that $e^{i\bk\cdot x}$ gets mapped to $-|\bk|^{2\alpha} e^{i\bk\cdot x}$ via standard Fourier transform calculations. Therefore, the operator in \eqref{levy5} has the same eigenfunctions as the Laplacian on $\mathbb T^2$ and the corresponding eigenvalues are the eigenvalues of the Laplacian raised to the power of $\alpha\in (0,1)$.


\subsection{Asymptotic derivation of leading order L\'evy flight search time} \label{appB}

In this section, we provide an asymptotic derivation of the leading order behavior of $\bar{u}_\eps$ in \eqref{ubarasympt}; i.e., the spatial average of $u_\eps(\bx)$ as $\eps \to 0^+$. We begin with the elliptic pseudo-differential equation \eqref{Au} for $u_\eps(\bx)$, where
$B_\eps(\bx_0)$ denotes the ball of radius $\eps$ centered at $\bx_0$. 

To simplify notation, we choose a coordinate system which identifies $\mathbb T^2$ with $[0,1)\times [0,1)$ so that $\bx_0 = (1/2,1/2)$, without loss of generality. This will ensure that the distance $d_{\mathbb T^2}(\cdot,\cdot)$ function on $\mathbb T^2$ satisfies $d_{\mathbb T^2}(\bx,\bx_0) = |\bx-\bx_0|$ for all $\bx \in \mathbb T^2 \cong [0,1)\times [0,1)$.

In the inner region, we let 
\begin{equation*}
	\bz = (\bx-\bx_0)/\eps \,, \qquad u_\eps(\bx_0 + \eps\bz) \sim U(\bz) \,.
\end{equation*}
In the inner variable $\bz$, we now show that $A_\alpha \sim -\eps^{-2\alpha}(-\Delta_\bz)^\alpha$ as $\eps \to 0^+$, where $ -\eps^{-2\alpha}(-\Delta_\bz)^\alpha$ is the fractional Laplacian on $\mathbb{R}^2$ with respect to the $\bz$ variable. To see this, we apply $\mathcal{A}_\alpha$ to $f(\eps^{-1}(\bx-\bx_0))$, which results in
\begin{multline} \label{frac1}
	\mathcal{A}_\alpha f(\eps^{-1}(\bx-\bx_0)) = \frac{4^\alpha \Gamma(1+\alpha)}{\pi |\Gamma(-\alpha)|} \\ \times   \int\limits_{[0,1]\times [0,1]} \! \sum_{\mathbf{m} \in \mathbb{Z}^2} \frac{f(\eps^{-1}(\by-\bx_0)) - f(\eps^{-1}(\bx-\bx_0))}{|\bx-(\by + \mathbf{m})|^{2+2\alpha}}  \, d\by\,.
\end{multline}
To obtain an expression in the form of the fractional Laplacian \eqref{fL} from \eqref{frac1}, we manipulate the denominator in the sum to obtain
\begin{widetext}
\begin{equation} \label{frac2}
	\mathcal{A}_\alpha f(\eps^{-1}(\bx-\bx_0)) = \frac{1}{\eps^{2+2\alpha}}\frac{4^\alpha \Gamma(1+\alpha)}{\pi |\Gamma(-\alpha)|} \int\limits_{[0,1]\times [0,1]} \! \sum_{\mathbf{m} \in \mathbb{Z}^2} \frac{f(\eps^{-1}(\by-\bx_0)) - f(\eps^{-1}(\bx-\bx_0))}{|\eps^{-1}(\bx-\bx_0)-(\eps^{-1}(\by-\bx_0) + \eps^{-1}\mathbf{m})|^{2+2\alpha}}  \, d\by\,.
\end{equation}
\end{widetext}
Substituting $\eps^{-1}(\by-\bx_0) \to \mathbf{Y}$ and $\eps^{-1}(\bx-\bx_0) \to \mathbf{X}$, \eqref{frac2} becomes
\begin{multline} \label{frac3}
	\mathcal{A}_\alpha f(\mathbf{X}) = \frac{1}{\eps^{2\alpha}}\frac{4^\alpha \Gamma(1+\alpha)}{\pi |\Gamma(-\alpha)|}\\ \times \int\limits_{R_{\frac{1}{\eps}}(\bx_0)} \! \sum_{\mathbf{m} \in \mathbb{Z}^2} \frac{f(\bY) - f(\bX)}{|\bX-(\bY + \eps^{-1}\mathbf{m})|^{2+2\alpha}}  \, d\by\,,
\end{multline}
where the region of integration $R_{\frac{1}{\eps}}(\bx_0)$ is the square of side length $\eps^{-1}$ centered at $\bx_0$. Now in the limit $\eps \to 0^+$, only the $\mathbf{m} = \mathbf{0}$ term in the sum of \eqref{frac3} contributes at a nonzero term, while the region of integration approaches $\mathbb{R}^2$, yielding
\begin{equation} \label{frac4}
	\mathcal{A}_\alpha f(\mathbf{X}) \sim \frac{1}{\eps^{2\alpha}}\frac{4^\alpha \Gamma(1+\alpha)}{\pi |\Gamma(-\alpha)|} \int\limits_{\mathbb{R}^2} \! \frac{f(\bY) - f(\bX)}{|\bX-\bY |^{2+2\alpha}}  \, d\bY\,.
\end{equation}
Comparing to \eqref{fL}, we find that the right-hand side of \eqref{frac4} is simply the fractional Laplacian with respect to the rescaled $\bX = \eps^{-1}(\bx-\bx_0))$ variable, scaled by a factor of $\eps^{-2\alpha}$. That is, $\mathcal{A}_\alpha \sim -\eps^{-2\alpha}(-\Delta_\bX)^\alpha$, as required.

We now expand $U \sim \eps^{2\alpha-2}U_0 + \eps^{2\alpha} U_1$, an expansion which we motivate below, so that the leading order term of the inner solution $U_0$ satisfies the radially symmetric exterior problem on $\mathbb{R}^2$:
\begin{subequations} \label{U}
	\begin{equation} \label{Ueq}
		-(-\Delta_\bz)^\alpha U_0 = 0\enspace {\rm on}\enspace \mathbb R^2 \setminus \overline {B_1(\mathbf{0})} \,, \quad U_0 = 0\enspace{\rm on}\enspace \overline{B_1(\mathbf{0})} \,;
	\end{equation}
	\begin{equation} \label{Ufar}
		U_0 \sim S_\alpha\left(-\frac{1}{|\bz|^{2-2\alpha}} + \chi_\alpha   \right) \ {\rm as}\ |\bz| \to \infty  \,,
	\end{equation}
\end{subequations}
where $S_\alpha$ is an $\mathcal{O}(1)$ constant to be found, $\chi_\alpha$ is a constant that depends on $\alpha$ and, in general, the geometry of the target. We show below that \eqref{U} may be reformulated as an integral equation on $B_1(\bzero)$ (i.e., the domain obtained by rescaling the target by $\eps^{-1}$ to $\mathcal{O}(1)$ size). For the special case here where $B_1(\bzero)$ is the unit ball, we then refer to \cite{kahane1981solution} for an explicit solution of this integral equation, which in turn yields an explicit expression for $\chi_\alpha$.

In the limit $\eps \to 0^+$ where the target size shrinks to zero, the exterior problem for $u_\eps$ in \eqref{integral equation} becomes the pseudo-differential equation in the punctured domain $\mathbb{T}^2\setminus~\lbrace \bx_0\rbrace$,
\begin{subequations} \label{ueps}
	\begin{equation} \label{uepseq}
		\mathcal{A}_\alpha u_\eps = -1  \,, \quad \bx \in \mathbb{T}^2 \setminus \lbrace \bx_0 \rbrace \,,
	\end{equation}
	with a prescribed local behavior near $\bx_0$, which is given by the far-field behavior of $U \sim \eps^{2\alpha-2}U_0$ in \eqref{Ufar},
	\begin{equation} \label{uepsloc}
		u_\eps \sim \eps^{2\alpha-2}S_\alpha\left( -\frac{\eps^{2-2\alpha}}{|\bx-\bx_0|^{2-2\alpha}} + \chi_\alpha    \right) \enspace \mbox{as} \enspace \bx \to \bx_0 \,
	\end{equation}
\end{subequations}
where $d_{\mathbb T^2}(\cdot,\cdot)$ is the distance on $\mathbb T^2$ with respect to the flat metric. We remark that, with $S_\alpha \sim \mO(1)$, \eqref{uepsloc} indicates that $u_\eps \sim \mO(1)$ in the outer region, which must be the case given that \eqref{uepseq} is independent of $\eps$. This requirement is what motivated the expansion for $U$ in the inner region.

Note also that \eqref{uepsloc} prescribes the singular structure of $u_\eps$ near $\bx_0$ as well as its regular part at $\bx_0$. Along with \eqref{uepseq}, \eqref{uepsloc} suggests that $u_\eps$ in the limit $\eps \to 0$ may be expressed in terms of the source-neutral Green's function $G_\alpha(\bx;\bx_0)$,
\begin{equation} \label{usol}
	u_\eps(\bx) \sim G_\alpha(\bx;\bx_0) + \bar{u}_\eps \,,
\end{equation}
where $\bar{u}_\eps$ is the spatial average of $u_\eps$, while $G_\alpha(\bx;\bx_0)$ satisfies
\begin{widetext}
\begin{subequations} \label{G}
	\begin{equation} \label{Geq}
		\mathcal{A}_\alpha G_\alpha = -1 + \delta(|\bx-\bx_0|) \,, \quad \bx \in \mathbb{T}^2 \setminus \lbrace \bx_0 \rbrace \,; \qquad \int_{\mathbb{T}^2} \! G_\alpha(\bx;\bx_0) \, d\bx = 0 \,;
	\end{equation}
	\begin{equation} \label{Gloc}
		G_\alpha(\bx;\bx_0) \sim -\frac{c_\alpha}{|\bx-\bx_0|^{2-2\alpha}} + R_\alpha(\bx_0;\bx_0) + \mathcal{O}(|\bx-\bx_0|)  \enspace \mbox{as} \enspace \bx \to \bx_0 \,; \qquad c_\alpha \equiv \frac{\Gamma(1-\alpha)}{4^\alpha\pi \Gamma(\alpha)} \,.
	\end{equation}
\end{subequations}
\end{widetext}
We remark that the pseudo-differential equation of \eqref{Geq} is consistent, since the right-hand side integrates to zero over $\mathbb{T}^2$. The integral condition of \eqref{Geq} is required to uniquely specify $G_\alpha$ since the constant function lies in the nullspace of $\mathcal{A}_\alpha$. In \eqref{Gloc}, the coefficient $c_\alpha$ is obtained simply by replacing $\alpha$ by $-\alpha$ in \eqref{fL}. In particular, the leading order singular term of $G_\alpha$ in \eqref{Gloc} is identical to the free space Green's function of \eqref{Gf} below. The regular part of the Green's function, denoted $R_\alpha(\bx_0;\bx_0)$ in \eqref{Gloc}, is given by \begin{equation} \label{Ralpha} R_\alpha(\bx_0;\bx_0) = \lim_{\bx \to \bx_0} G_\alpha(\bx;\bx_0) + \frac{c_\alpha}{|\bx-\bx_0|^{2-2\alpha}} \,.\end{equation}

In fact, if we set $\rho, \tilde\rho\in C^\infty_c(\mathbb R)$ satisfying $\rho(t) = 1$ for $t\in (-1/8, 1/8)$ and $\supp(\rho) \subset (-1/4, 1/4)$ and $\tilde\rho (t) = 1$ for $t\in (-1/4, 1/4)$ and $\supp(\tilde\rho) \subset (-1/2,1/2)$, we can write
\begin{eqnarray*}
\left(\mathcal{A}_\alpha \frac{c_\alpha \rho(\cdot,\bx_0)}{|\cdot-\bx_0|^{2-2\alpha}} \right)&=& \tilde\rho(\cdot, \bx_0)\left(\mathcal{A}_\alpha \frac{c_\alpha \rho(\cdot,\bx_0)}{|\cdot-\bx_0|^{2-2\alpha}} \right) \\&+& \left(1- \tilde\rho(\cdot, \bx_0)\right)\left(\mathcal{A}_\alpha \frac{c_\alpha \rho(\cdot,\bx_0)}{|\cdot-\bx_0|^{2-2\alpha}} \right)\,,
\end{eqnarray*}
where $\rho(\bx, \bx_0) = \rho(|\bx-\bx_0|^2)$ and same for $\tilde\rho(\bx,\bx_0)$. Observe that since $\left(1- \tilde\rho(\cdot, \bx_0)\right)$ and $\rho(\cdot,\bx_0)$ has disjoint support, the second term is infinitely smooth by the fact that $\mathcal A_\alpha$ is given by an integral kernel which is smooth away from the diagonal. For the first term, we can directly compute using normal coordinates centered at $\bx_0$ to obtain
$$-\left(\mathcal{A}_\alpha \frac{c_\alpha \rho(\cdot,\bx_0)}{|\cdot-\bx_0|^{2-2\alpha}} \right)\left(\bx\right) = \delta(|\bx_0-\bx|) + C^\infty(\mathbb T^2).$$
We may then write $G_\alpha(\bx;\bx_0) = -\frac{c_\alpha \rho(\bx,\bx_0)}{|\bx-\bx_0|^{2-2\alpha}}+ R_\alpha(\bx;\bx_0)$ where $R_\alpha(\cdot;\bx_0) : \mathbb T^2 \to \mathbb R$ satisfies $\mathcal{A}_\alpha R_\alpha (\cdot; \bx_0) = -1 + C^\infty(\mathbb T^2)$. It can then be shown (see Ch. 10 of \cite{taylor2013partial}) that $R_\alpha(\bx;\bx_0)$ is infinitely smooth on $\mathbb T^2$.

To determine $S$ and $\bar{u}_\eps$, we perform a leading order matching of the local behavior of $u_\eps$ in \eqref{usol} to the required singularity structure of \eqref{uepsloc}. That is, from \eqref{usol} and \eqref{Gloc}, we have that as $\bx \to \bx_0$,
\begin{multline} \label{match}
	-\frac{c_\alpha}{|\bx-\bx_0|^{2-2\alpha}} + R_\alpha(\bx_0;\bx_0) + \bar{u}_\eps  = \\ \eps^{2\alpha-2}S_\alpha\left( -\frac{\eps^{2-2\alpha}}{|\bx-\bx_0|^{2-2\alpha}} + \chi_\alpha    \right) \,.
\end{multline}
Matching $\mathcal{O}(|\bx-\bx_0|^{2\alpha-2})$  and $\mathcal{O}(1)$, we arrive at $S_\alpha = c_\alpha$, and the GMFPT of $u_\eps$,
\begin{equation} \label{ubareps}
	\bar{u}_\eps =  \eps^{2\alpha-2}\frac{\Gamma(1-\alpha)}{4^\alpha\pi \Gamma(\alpha)}\chi_\alpha - R_\alpha(\bx_0;\bx_0) \,.
\end{equation}
We note that the leading order term \eqref{ubareps} depends on the geometry of target through the constant $\chi_\alpha$ to be computed below. This is in contrast to the 2-D narrow capture problem with Brownian motion, where target geometry effects enter only at the $\mathcal{O}(1)$ correction term. Global geometric properties of the search domain are contained in the $\mathcal{O}(1)$ correction term of \eqref{ubareps} through the regular part of the Green's function \eqref{Ralpha}. \blue{We remark also that the dependence of search time on the starting location $\bx$ enters only at $\mO(1)$ through $G_\alpha(\bx;\bx_0)$ in \eqref{usol}, which is subdominant to the $\mO(\eps^{2\alpha-2})$ constant term $\bar{u}_\eps$. Thus, in the $\eps\to 0^+$ limit considered here, the better search strategy will always be the one for which $\alpha$ is closer to the Brownian limit of 1, regardless of from where the search begins. } 

To determine $\chi_\alpha$, we require the solution to \eqref{U}. Letting $U_0(\bz) = S_\alpha u(\bz)$ in \eqref{U}, we have for $u(\bz)$,
\begin{subequations} \label{R2}
	\begin{equation} \label{R2eq}
		-(-\Delta)^\alpha u = 0\enspace {\rm on}\enspace \mathbb R^2 \setminus \overline {B_1(\mathbf{0})} \,, \quad u = 0\enspace {\rm on}\enspace \overline{B_1(\mathbf{0})} \,.
	\end{equation}
	\begin{equation} \label{R2far}
		u \sim -\frac{1}{|\bz|^{2-2\alpha}} + \chi_\alpha    \enspace {\rm as} \enspace |\bz| \to \infty  \,,
	\end{equation}
\end{subequations}
where $-(-\Delta)^\alpha$ is with respect to the $\bz$ variable. Note that we have normalized the coefficient of $|\bz|^{2\alpha-2}$ in the far-field behavior \eqref{R2far}, which leaves $\chi_\alpha$ as the parameter to be determined.

We next let $$v(\bz) = u(\bz)-\chi_\alpha \,,$$ so that $v$ satisfies 
\begin{subequations} \label{v}
	\begin{equation} \label{veq}
		-(-\Delta)^\alpha v = 0\enspace {\rm on}\enspace \mathbb R^2 \setminus \overline {B_1(\mathbf{0})} \,, \quad v = -\chi_\alpha\enspace {\rm on}\enspace \overline{B_1(\mathbf{0})} \,;
	\end{equation}
	\begin{equation} \label{vfar}
		v \sim -\frac{1}{|\bz|^{2-2\alpha}}    \enspace {\rm as} \enspace |\bz| \to \infty  \,,
	\end{equation}
\end{subequations}
The exterior problem \eqref{v} for $v(\bz)$ may be reformulated as the following problem over all of $\mathbb{R}^2$ without boundary,
\begin{subequations} \label{vR2}
	\begin{equation} \label{vR2eq}
		-(-\Delta)^\alpha v = \frac{1}{c_\alpha}f(\bz) \enspace {\rm on}\enspace \mathbb R^2 \,; \quad f(\bz) \equiv 0 \enspace {\rm on}\enspace \mathbb{R}^2\setminus\overline{B_1(\mathbf{0})} \,;
	\end{equation}
	\begin{equation} \label{vR2far}
	v \equiv -\chi_\alpha \enspace \mbox{on} \enspace \overline{B_1(\mathbf{0})}  \,; \quad 	v \sim -\frac{1}{|\bz|^{2-2\alpha}}    \enspace {\rm as} \enspace |\bz| \to \infty  \,,
	\end{equation}
\end{subequations}
where $c_\alpha$ is the constant given in \eqref{Gloc}, and $f(\bz)$ is an unknown function to be found by imposing the first condition in \eqref{vR2far}. We now derive this integral equation for $f(\bz)$. 

First, the free space Green's function $G_f(\bz;\bzero)$ with source centered at the origin satisfying
\begin{subequations} \label{Gf}
	\begin{equation} \label{Gfeq}
		-(-\Delta)^\alpha G_f = \delta(\bz) \enspace {\rm on}\enspace \mathbb R^2 \,;
	\end{equation}
	\begin{equation} \label{Gffar}
		G_f(\bz;\bzero) \to 0   \enspace {\rm as} \enspace |\bz| \to \infty  \,,
	\end{equation}
\end{subequations}
is given by
\begin{equation} \label{Gfsol}
	G_f(\bz;\bzero) = -c_\alpha \frac{1}{|\bz|^{2-2\alpha}} \,,
\end{equation}
where $c_\alpha$ is given in \eqref{Gloc}. Then the solution $v(\bz)$ of \eqref{vR2} may be written as a convolution of the right-hand side of \eqref{vR2} with $G_f(\bz;\bzero)$ of \eqref{Gfsol}, which yields
\begin{equation} \label{vsol}
	v(\bz) = -\int_{B_1(\bzero)}\! \frac{f(\by)}{|\bz-\by|^{2-2\alpha}} \, d\by \,,
\end{equation}
where the region of integration in \eqref{vsol} is only over $B_1(\bzero)$ because $f(\bz)$ is compactly supported in $B_1(\bzero)$ (see \eqref{vR2eq}). To impose the normalizing condition on $v$ given in the far-field condition\eqref{vR2far}, we expand the kernel as $|\bz| \to \infty$ in \eqref{vsol} as $|\bz-\by|^{2\alpha-2} \sim |\bz|^{2\alpha-2} + \mathcal{O}(|\bz|^{2\alpha-3})$, noting that $|\by| \leq 1$ by virtue of the region of integration. Substituting this leading order expansion into \eqref{vsol} and comparing to the required far-field behavior of $v(\bz)$ in \eqref{vR2far}, we obtain the normalizing condition for $f(\bz)$,
\begin{subequations} \label{vint}
	\begin{equation} \label{vintnormal}
		\int_{B_1(\bzero)} \! f(\bz) \, d\bz = 1 \,.
	\end{equation}
	Next, we require that $v(\bz) \equiv -\chi_\alpha$ for $\bz \in B_1(\bzero)$, yielding
	\begin{equation} \label{vinteq}
		\int_{B_1(\bzero)}\! \frac{f(\by)}{|\bz-\by|^{2-2\alpha}} \, d\by = \chi_\alpha \,, \quad \bz \in \overline{B_1(\bzero)} \,.
	\end{equation}
\end{subequations}

The integral equation \eqref{vinteq} together with the normalizing condition \eqref{vintnormal} are to be solved simultaneously for $f(\bz)$ and $\chi_\alpha$. To determine an explicit solution to \eqref{vint}, we appeal to the result of \cite{kahane1981solution} (see Theorem 3.1, and, in particular, (3.37)), which states that for $0<\beta<1$ and any $\bz$ such that $|\bz|\leq 1$,
\begin{subequations} \label{inteqsolall}
	\begin{equation} \label{inteqsol}
		\int_{B_1(\bzero)} \! \frac{1}{|\bz-\by|^{2\beta}} \frac{1}{(1-|\by|^2)^{1-\beta}} d\by = \frac{\pi^2}{\sin(\beta\pi)} \,.
	\end{equation}
	Identifying $\beta$ in \eqref{inteqsol} with $1-\alpha$, and noting that
	\begin{equation}
		\int_{B_1(\bzero)} \! \frac{1}{(1-|\by|^2)^{1-\beta}} \, d\by = \frac{\pi}{\beta} \,,
	\end{equation}
\end{subequations}
we find from \eqref{inteqsolall} that
\begin{multline}
	\int_{B_1(\bzero)} \! \frac{1}{|\bz-\by|^{2-2\alpha}} \frac{(1-\alpha)/\pi}{(1-|\by|^2)^{\alpha}} d\by = \\ \frac{1-\alpha}{\pi}\frac{\pi^2}{\sin((1-\alpha)\pi)} \,.
\end{multline}
Comparing \eqref{inteqsol} with \eqref{vint}, we find that the solution to \eqref{vint} is given by
\begin{equation} \label{vintsol}
	f(\bz) = \frac{1-\alpha}{\pi(1-|\bz|^2)^{\alpha}} \,, \qquad  \chi_\alpha = \frac{\pi(1-\alpha)}{\sin((1-\alpha)\pi)} \,.
\end{equation}
Substituting $\chi_\alpha$ from \eqref{vintsol} into the expression for the GMFPT $\bar{u}_\eps$ in \eqref{ubareps}, with $0<\alpha<1$, we arrive at $\bar{u}_\eps$ as given in \eqref{ubarasympt}.

For the GMFPT $\bar{u}_\eps^{(L)}$ on the flat 2-D torus of side length $L$ with a circular target of radius $\eps$, a simple rescaling shows that $\bar{u}_\eps^{(L)} = L^2 \bar{u}_\eps$ where $\bar{u}_\eps$ is the GMFPT on the flat torus of unit side length given in \eqref{ubarasympt}. This scaling was obtained empirically through particle simulations in \cite{levernier2018universal}. In contrast, our systematic approach via a asymptotic analysis resulting in \eqref{ubarasympt} yields not only the $\alpha$-dependent prefactor in the leading order $\mO(\eps^{2\alpha-2})$ term, but also the geometry-dependent $\mO(1)$ correction term. Neither of these terms are given in \cite{levernier2018universal}, which focuses on much broader class of random walks, \blue{including L\'evy walks where particle velocity is fixed in contrast to the possibility of unbounded velocities allowed by L\'evy flights (see also \cite{dybiec2017levy})}. In the next section, we outline the matched asymptotic method for the flat 3-D torus as well as closed 2-D manifolds.

\subsection{Matched asymptotic method for the (flat) 3-D torus and 2-D manifolds} \label{appC}

We now give a brief description of how the analysis of \S \ref{appB} can be extended to the flat 3-D torus as well as 2-D closed manifolds. For the unit 3-D torus $\mathbb{T}^3 = [0,1)^3$ with a spherical target of radius $\eps$ centered at $\bx_0 = (1/2,1/2, 1/2)$, we have the elliptic problem for the MFPT $u_\eps(\bx)$ (in analogy with \eqref{integral equation}),
\begin{subequations} \label{Au3app} \begin{equation} \label{integral equation app3}
		\mathcal{A}_\alpha u_\eps = -1\enspace {\rm on}\enspace \mathbb T^3\setminus \overline {B_\eps(\bx_0)},\ \ u_\eps = 0\enspace {\rm on }\enspace \overline{B_\eps(\bx_0)} \,,
	\end{equation}
	where $\mathcal{A}_\alpha$ for $0<\alpha<1$ is the fractional Laplacian of order $\alpha$ on $\mathbb{T}^3$ given by \begin{equation}  \label{fraclapapp3}
		\mathcal{A}_\alpha f(\bx) :=  \frac{4^\alpha \Gamma(3/2+\alpha)}{\pi^{3/2} |\Gamma(-\alpha)|} \\ \times\int\limits_{\mathbb{T}^3} \! \sum_{\mathbf{m} \in \mathbb{Z}^3} \frac{f(\by) - f(\bx)}{|\bx-(\by + \mathbf{m})|^{3+2\alpha}} \, d\by\,.
	\end{equation}
\end{subequations}
This fact can be seen by running the same argument as that at the end of \S \ref{appA} for the fractional Laplacian on the flat 2-D torus.

To analyze $u_\eps(\bx)$ near $\bx_0$, we change variables to $\bx = \bx_0 + \eps\bz$ and consider the inner problem on $\mathbb{R}^3$,
\begin{subequations} \label{U3}
	\begin{equation} \label{Ueq3}
		-(-\Delta_\bz)^\alpha U_0 = 0\enspace {\rm on}\enspace \mathbb R^3 \setminus \overline {B_1(\mathbf{0})} \,, \quad U_0 = 0\enspace{\rm on}\enspace \overline{B_1(\mathbf{0})} \,;
	\end{equation}
	\begin{equation} \label{Ufar3}
		U_0 \sim S_\alpha\left(-\frac{1}{|\bz|^{3-2\alpha}} + \chi_\alpha   \right) \ {\rm as}\ |\bz| \to \infty  \,,
	\end{equation}
\end{subequations}
where $-(-\Delta_\bz)^\alpha$ in \eqref{Ueq3} is the 3-D fractional Laplacian given by
\begin{equation}
	-(-\Delta_\bz)^\alpha f(\bx) = \frac{1}{\eps^{2\alpha}}\frac{4^\alpha \Gamma(3/2+\alpha)}{\pi^{3/2} |\Gamma(-\alpha)|} \int\limits_{\mathbb{R}^2} \! \frac{f(\by) - f(\bx)}{|\bx-\by |^{3+2\alpha}}  \, d\by\,.
\end{equation}
Following the same analysis that resulted in \eqref{ubareps}, we obtain that the GMFPT of a search of a small circular target of radius $\eps$ in $\mathbb{T}^3$ is given by
\begin{equation} \label{ubareps3}
	\bar{u}_\eps =  \eps^{2\alpha-3}\frac{\Gamma(3/2-\alpha)}{4^\alpha\pi^{3/2} \Gamma(\alpha)}\chi_\alpha - R_\alpha(\bx_0;\bx_0) \,,
\end{equation}
where $R_\alpha(\bx_0;\bx_0)$ is the regular part of the Green's function satisfying
\begin{widetext}
	\begin{subequations} \label{G3}
		\begin{equation} \label{Geq3}
			\mathcal{A}_\alpha G_\alpha = -1 + \delta(|\bx-\bx_0|) \,, \quad \bx \in \mathbb{T}^3 \setminus \lbrace \bx_0 \rbrace \,; \qquad \int_{\mathbb{T}^3} \! G_\alpha(\bx;\bx_0) \, d\bx = 0 \,;
		\end{equation}
		\begin{equation} \label{Gloc3}
			G_\alpha(\bx;\bx_0) \sim -\frac{c_\alpha}{|\bx-\bx_0|^{3-2\alpha}} + R_\alpha(\bx_0;\bx_0) + \mathcal{O}(|\bx-\bx_0|)  \enspace \mbox{as} \enspace \bx \to \bx_0 \,; \qquad c_\alpha \equiv \frac{\Gamma(3/2-\alpha)}{4^\alpha\pi^{3/2} \Gamma(\alpha)} \,.
		\end{equation}
	\end{subequations}
\end{widetext}
To compute $\chi_\alpha$, we follow the analysis leading to \eqref{vint} to recast \eqref{U3} as an integral equation inside the 3-D unit ball $B_1(\bzero)$,
\begin{subequations} \label{vint3}
	\begin{equation} \label{vinteq3}
		\int_{B_1(\bzero)}\! \frac{f(\by)}{|\bz-\by|^{3-2\alpha}} \, d\by = \chi_\alpha \,, \quad \bz \in \overline{B_1(\bzero)} \,.
	\end{equation}
	subject to the normalization condition	
	\begin{equation} \label{vintnormal3}
		\int_{B_1(\bzero)} \! f(\bz) \, d\bz = 1 \,.
	\end{equation}
\end{subequations}

The solution to \eqref{vinteq3}, as given by \cite{kahane1983solution}, is
\begin{subequations} \label{f3}
\begin{equation} \label{f3sol}
	f(\by) = \frac{\sin((1-\alpha)\pi}{\pi^{5/2}}\frac{\Gamma(3/2-\alpha)}{\Gamma(1-\beta)}\chi_\alpha I\frac{1}{(1-|\by|^2)^\alpha}  \,;
\end{equation}
\begin{equation} \label{f3I}
	I \equiv -\frac{\alpha\tB(1/2, -\alpha)}{2(1/2-\alpha)} \,,
\end{equation}
\end{subequations}
where the beta function in \eqref{f3I} is defined by $\tB(z_1, z_2) \equiv \Gamma(z_1)\Gamma(z_2)/\Gamma(z_1+z_2)$ is the beta function. Imposing the integral condition \eqref{vintnormal3} on $f(\by)$ in \eqref{f3} yields $\chi_\alpha$, given by
\begin{equation} \label{chialpha3}
	\chi_\alpha = \frac{2\sqrt{\pi}(3/2-\alpha)(1/2-\alpha)\Gamma(1/2-\alpha)}{\alpha |\Gamma(-\alpha)|\sin((1-\alpha)\pi)} \,.
\end{equation}
We note that the quantity $(1/2-\alpha)\Gamma(1/2-\alpha) > 0$ for $\alpha \in (0,1)$, so $\chi_\alpha> 0$. Together with \eqref{ubareps3}, the global MFPT is given to $\mathcal{O}(1)$ by
\begin{equation} \label{ubareps3full}
	\bar{u}_\eps =  \eps^{2\alpha-3}\frac{2}{4^\alpha\pi\alpha}\frac{\left\lbrack(1/2-\alpha)\Gamma(1/2-\alpha)\right\rbrack^2}{\Gamma(\alpha)|\Gamma(-\alpha)|\sin((1-\alpha)\pi)} - R_\alpha(\bx_0;\bx_0) \,.
\end{equation}
Since $\alpha \in (0,1)$, the leading order scaling of $\bar{u}_\eps$ in \eqref{ubareps3full} is asymptotically larger than the $\mathcal{O}(\eps^{-1})$ search time of a Brownian particle in a confined 3-D volume \cite{cheviakov2010asymptotic}.

The analysis on a closed compact 2-D manifold $M$ of unit area and which is independent of the target size $\epsilon$ can proceed in a manner similar to the matched asymptotic method of \S \ref{appB}, with the operator $\mathcal{A}_\alpha$ replaced by that given in the Introduction section of \cite{applebaum2000isotropic} (with the parameter $a = 0$, signifying pure L\'evy flight). This operator (denoted $\mathscr{A}$ in \cite{applebaum2000isotropic}) assumes that a jump from a point $p \in M$ occurs along any one of the geodesics emanating from $p$, and that the length of this jump is measured along that geodesic curve. A circular target $B_\eps(p_0)$ of radius $0 < \varepsilon \ll 1$ centered about $p_0 \in M$ is then taken to be the set of all points on $M$ whose distance from $p_0$ is less than or equal to  $\eps$. 

In the $\mathcal{O}(\eps)$ region centered about $p_0$ and in the appropriate Riemannian normal coordinates (see \cite{tzou2020analysis} for a similar matched asymptotic analysis involving the Laplace-Beltrami operator), it can be shown that $\mathscr{A}$ to leading order in $\eps$ reduces to the (flat) fractional Laplacian on $\mathbb{R}^2$. In these coordinates, the analysis of $u_\eps$ in the inner region proceeds exactly as for the 2-D torus in \S \ref{appB}. Analysis of $u_\eps$ in the outer region away from $p_0$ also remains largely unchanged with the main exception being that the Green's function would involve the operator $\mathscr{A}$ of \cite{applebaum2000isotropic} instead of $\mathcal{A}_\alpha$ in \S \ref{numerics}. 

As a result, we predict that the global MFPT on a closed 2-D manifold of unit area will possess the same leading order behavior \eqref{ubarasympt}. However, depending on $\alpha$ and the curvature of $M$ at the center of the target, weaker algebraic and/or logarithmic singularities in $\eps$ may follow the leading $\mathcal{O}(\eps^{2\alpha-2})$ term. Moreover, computation of the regular part of the Green's function for the $\mathcal{O}(1)$ correction term may be challenging for general manifolds due to the difficulty of computing the operator $\mathscr{A}$ itself. Further complications (and interesting behavior of $u_\eps$) may arise if $M$ contains points connected to $p_0$ by a family of geodesics that can be parameterized by a continuous parameter (such pairs of points are absent on the flat torus).

\section{Discussion}

Through Monte Carlo simulations, direct numerical solutions and asymptotic analysis of the limiting nonlocal exterior problem \eqref{Au}, we have shown that the average search time of a L\'evy flight with tail index $0<\alpha<1$ on the flat torus $\mathbb{T}^2$ with a small circular target of radius $0 < \eps \ll 1$ scales as $\mathcal{O}(\eps^{2\alpha-2})$. In addition, our asymptotic analysis has yielded the $\mathcal{O}(1)$ correction term to search time, along with insights into how target geometry and search domain geometry and size impact search times. 

By comparing to average search times of the Brownian walk on the same domain, which obey the well-known $\mathcal{O}(|\log\eps|)$ scaling, we have provided a concrete counterexample to the L\'evy flight foraging hypothesis. We emphasize that our comparison is limited only to the narrow escape framework, and is not a general statement on the superiority of Brownian over L\'evy search strategies. One possible avenue may be to assess whether a search strategy based on a combination of Brownian motion and L\'evy flight (e.g., \cite{palyulin2016search}) can be optimized to be faster than pure Brownian search in the narrow escape framework. For this, the result of \cite{applebaum2000isotropic} may be useful, where an infinitesimal generator is obtained that governs the continuous limit of a L\'evy process in which Brownian motion is ``interlaced'' with long jumps. 


While we presented results only for small target sizes $\eps$, we note that L\'evy search times exceeded Brownian search times for all $\eps < 0.5$. Within this first-passage time framework, it may be insightful to seek possible scenarios in which the L\'evy search strategy is superior.

We now discuss some avenues for future work, several of which are projects currently in progress. While a single target $\mathbb{T}^2$ is a very simple domain on which to perform this comparison, it would be interesting to consider more complex domains. For example, a finite domain with reflecting boundaries containing perhaps small reflecting obstacles may present challenges from both a modeling and analytic perspective. From a particle simulations perspective, reflective domains and obstacles would require computing trajectories of flights that undergo reflections. From the pseudo-differential equation perspective, one must formulate an analog to the $\mathcal{A}_\alpha$ operator of \eqref{fraclap}; this new operator must account for all possible paths between all pairs of points, including those that reflect off boundaries and obstacles.

A domain featuring non-constant curvature would also present computational challenges - geodesics would need to be computed for both the Monte Carlo algorithm as well as the finite difference method for discretizing the corresponding infinitesimal generator (see \cite{chaubet2022geodesic}). This would add to the already significant computational cost. The sphere, on the other hand, has simple geodesics and may be a good candidate for a follow-up study, especially considering the interesting predictions made in \cite{chaubet2022geodesic} regarding the possible optimality of starting the search from the point antipodal to the center of the target (see Theorem 1.1 part (iii)).

Another domain feature that we have not considered is the inclusion of more than one target, one or more of which may be of non-circular shape. The multiple-target problem has been considered at length for Brownian motion on flat 2- and 3-dimensional geometries using hybrid asymptotic-numerical methods see (e.g., \cite{delgado2015conditional, lindsay2017optimization, pillay2010asymptotic, cheviakov2010asymptotic, coombs2009diffusion, bressloff2013stochastic, bressloff2021asymptotic2,holcman2015stochastic} and the references therein). Several numerical optimization studies have been done to find optimal target arrangements that minimize the spatial average of the stopping time (e.g., \cite{cheviakov2013narrow, cheviakov2012mathematical, ridgway2018iterative, cheviakov2011optimizing}). The inclusion of more than one target also gives rise to the question of splitting probabilities (see e.g., \cite{kurella2015asymptotic, schuss2009theory, chevalier2010first}) and shielding effects \cite{kurella2015asymptotic}, and how they compare to their Brownian counterparts. 

One useful aspect of such a hybrid methods is their ability to capture the higher order correction terms of the GMFPT, which encode effects of target locations/configurations. This can be accomplished by performing a higher order matching in the asymptotic solution for $u_\eps(\bx)$ in Section B, and computing the regular part of the Green's function $G_\alpha$ in \eqref{G}. The greater ease of solving this $\eps$-independent problem without singular features has made possible the computational optimization studies referenced above. An analogous hybrid analytic-numerical theory for the fractional Laplacian on $\mathbb{T}^2$ or the more general operator on Riemannian manifolds would open various avenues of research. A similar method was recently developed for the Laplacian on Riemannian 2-manifolds using techniques in microlocal analysis \cite{tzou2019spot, tzou2020analysis}, which allowed for predictions of localized spot dynamics in reaction-diffusion systems on manifolds. 

Numerical results suggest that the variance of the stopping time may have a larger scaling with $\mathcal{O}(1/\eps)$ than the mean. The large variance suggests that the mean of the stopping time may not be particularly informative of the probability distribution of stopping times. Asymptotic computation of the variance of the MFPT for the narrow escape problem has been done in, e.g., \cite{kurella2015asymptotic, lindsay2016hybrid} for Brownian motion. To capture \textit{all} moments of the probability distribution, however, would require analysis of the diffusion equation. In \cite{lindsay2016hybrid, bressloff2021asymptotic}, a Laplace transform in the time variable was employed to transform the problem to an elliptic boundary value problem, on which the hybrid asymptotic-numerical tools of \cite{ward1993summing} could be applied before transforming back. A analogous approach may be possible to characterize the full distribution of stopping times of a L\'evy flight on Riemannian manifolds.

Finally, for Brownian motion, the problem of finding target configurations that optimize GMFPT is closely related to the problem of finding stable equilibrium configurations of localized spots in the Schnakenberg reaction-diffusion system (cf. \cite{tzou2017stability} and \cite{cheviakov2011optimizing}). Furthermore, the question of whether a \textit{mobile} target leads to lower GMFPT has been found to be closely related to a certain instability of the aforementioned localized spot equilibria \cite{tzou2014first, tzou2015mean, xie2017moving}. It would be interesting to explore whether these relationships still hold when Brownian motion is replaced by L\'evy flights.

\begin{acknowledgments}
JCT was supported by Australian Research Council Discovery Project DP220101808.
\end{acknowledgments}

\bibliography{rp}

\end{document}